\documentclass[twocolumn,twoside,slac_two]{revtex4}

\usepackage{graphicx}
\usepackage{fancyhdr}

\usepackage{amssymb}

\newcommand{\tbeta}{\tilde{\beta}}

\newcommand{\alp}{\alpha}     
     
\newcommand{\eps}{\epsilon}

\newcommand{\cA}{{\cal A}}    
\newcommand{\cC}{{\cal C}}

\begin{document}

\title{Extracting New Physics from the CMB}

%

\author{Brian Greene}
\affiliation{Institute for Strings, Cosmology and Astroparticle
Physics \\ Department of Physics, Columbia University, New York, NY
10027, USA}
\author{Koenraad Schalm}
\affiliation{Institute for Strings, Cosmology and Astroparticle
Physics \\ Department of Physics, Columbia University, New York, NY
10027, USA}
\author{Jan Pieter van der Schaar}
\affiliation{Institute for Strings, Cosmology and Astroparticle
Physics \\ Department of Physics, Columbia University, New York, NY
10027, USA}

\author{Gary Shiu}
\affiliation{Department of Physics, University of Wisconsin, Madison,
WI 53706, USA}

\begin{abstract}
We review how initial state effects generically yield an oscillatory
component in the primordial power spectrum of inflationary density
perturbations. These oscillatory corrections parametrize unknown new
physics at a scale $M$ and are potentially observable if the ratio
$H_{infl}/M$ is sufficiently large. We clarify to what extent present and
future CMB data analysis can distinguish between the different
proposals for initial state corrections.
\end{abstract}

\maketitle

\thispagestyle{fancy}


\section{Transplanckian Physics --- or, can Quantum Gravity be seen
in the sky?}

There is much about the early universe that remains beyond the
reach of today's most refined theoretical tools. Among the many
intertwined and as yet poorly understood issues are the nature and
resolution of the big bang singularity, the correct form of
physical laws in the extreme environment of the Planck era, and
the full specification of initial conditions for all physical
degrees of freedom. Even without answers to these questions,
however, cosmology has made great strides in recent years. This is
at least partly due to the happy fact that inflationary cosmology
--- viewed as an effective theory that describes the dynamics of
the universe at sufficiently ``late" times --- has a tendency to
suppress dependence on unknown physics of the very early universe.

Nevertheless, there are features of inflationary cosmology that
retain a memory of conditions and dynamics of the very early
universe, and a growing cadre of researchers have, in recent years,
tried to exploit this to provide a cosmological window on the Planck
era --- a body of work that is often referred to as
\emph{transplanckian physics} \cite{Brandenberger, egks, MNPH,
Others, Kaloper, Burgess, BEFT, generic-NPH, Porrati, back2, Pheno}.
In this note, we emphasize one such approach: seeking transplanckian
signatures in the cosmic microwave background radiation (CMB).

In the following sections, we will review potential transplanckian
signatures, emphasizing observational consequences over technical
details (which are covered in the references we cite). But first,
we give a quick sketch of the essential physics.

The standard, and highly successful, calculations of the CMB power
spectrum\footnote{For a recent theoretical review of CMB physics,
also covering some of the issues raised here, see
\cite{Giovannini}.}, rely on two essential assumptions:

(1) the standard dynamics of flat spacetime quantum field theory is
applicable on arbitrarily short scales (and hence arbitrarily high
energies) and

(2) the standard boundary conditions used in flat spacetime
quantum field theory are applicable when a mode's wavelength is
sufficiently small (the intuition here is that the smaller a
mode's physical wavelength --- the more blueshift its corresponding
comoving mode experiences --- the less sensitive it is to any
background spacetime curvature).

Transplanckian studies of the CMB challenge one or both of these
assumptions, and the literature is now replete with many specific
alternative proposals --- alternative dynamics and/or alternative
boundary conditions. We have argued \cite{egks,generic-NPH} that a
generic signature of such proposals is a new \emph{oscillatory}
feature overlaid on the usual primordially power spectrum. It is
straightforward to understand why: regardless of the primordial
dynamics and primordial boundary conditions, at sufficiently late
times (for any given mode) the successful standard dynamics ---
essentially Einstein's equations (or Einstein's equations with
couplings to a scalar field theory) --- must be the controlling
framework. At this late time, we can summarize the unknown
primordial dynamics and primordial boundary conditions through the
specification of boundary conditions to the Einstein equations. Of
course, an arbitrary choice of boundary conditions will result in
arbitrary results. The data, however, winnow the possibilities
since the boundary conditions must yield results that do not
differ significantly from the observed scale invariance. This
suggests two physically well-motivated classes of boundary
conditions.

(a) For each comoving mode $k$, set boundary conditions when the
physical momentum $k/a(t)$ is redshifted to a physical cutoff
scale $M_{cutoff}$ (e.g. the string scale in string theory), and
choose the boundary conditions to be nominally scale invariant by
making them depend only on the physical scale $k/a(t_k) =
M_{cutoff}$, or

(b) At a chosen time $t_{cutoff}$ (essentially, the earliest time
for which we can trust standard general relativistic dynamics), set
the boundary conditions for all modes $k$ on this equal time
hypersurface, and choose these boundary conditions to include
one-loop corrections to the standard (scale invariant) flat
spacetime boundary values.

In either case, the modified boundary conditions on each mode
amount to a Bogoliubov rotation of positive and negative frequency
components of that mode (relative to the standard vacuum choice).
Since the power spectrum is proportional to the square of a given
mode's amplitude, this rotation leads to a positive/negative mode
mixing, yielding the oscillatory behaviour referred to above. In
case (a), though, the argument of the oscillatory terms will
depend on $H_{inf}/(k/a(t_k)) = H_{inf}/M_{cutoff}$, which is
constant in de Sitter space, and hence truly oscillatory behaviour
only occurs in the physically relevant case of backgrounds with
non-constant Hubble parameter. In case (b), the oscillatory
behaviour is already present in de Sitter space as
$k/a(t_{cutoff})$ is explicitly $k$ dependent.

Thus, our main conclusion is that if transplanckian physics is
observable in the CMB --- admittedly a significant ``if'' as we need the
amplitude of the transplanckian contribution to be sufficiently
large --- then a prime signature to look for is an oscillatory
component to the primordial power spectrum.

In what follows, we spell this out in somewhat greater detail,
focusing on the choice of initial conditions in the context of
effective field theory --- a framework we feel to be both conservative
and reliable, but sufficiently rich to allow the calculation of the
form of oscillatory power spectrum component. We compare the results
found in the two cases (a) and (b), above, and note significant
qualitative differences.

\section{Initial state effects in the CMB and their relation to new
physics}

The initial state problem can be turned into an opportunity to
probe new high energy, or `transplanckian', physics, {\it if}
initial state selection proves to be related to physics at the
high energy scale, typically corresponding to the string or Planck
scale. Although many proposals have been put forward suggesting
such a link, they typically rely on highly particular models of
Planck scale physics that are predominantly ad hoc and contain
specifics whose justification can be questioned
\cite{Brandenberger, egks, MNPH, Others}. However,  the
{\em generic} features of Planck scale physics ought to be
describable by an effective field theory \cite{Kaloper}. Inspired by
\cite{Burgess}, \cite{BEFT} showed how initial conditions are
translated into the language of effective field theory (EFT)
through the introduction of a boundary action. This spacelike
boundary action is located at an initial time surface $t_0$ where
the initial conditions are set for {\it all} the bulk modes.
Primarily for phenomenological reasons\footnote{One can show that
the
  Bunch-Davies state is special from the boundary effective action
  point of view as well; see \cite{BEFT}.}
the boundary action is chosen to describe small corrections to the
Bunch-Davies (BD) state. The BD state corresponds to a specific
choice for a (relevant) operator on the boundary. The effect of
unknown Planck scale physics on the initial conditions is
parametrized in terms of irrelevant boundary operators. Their
presence induces small corrections to the BD state suppressed by
powers of the ratio of the physical momentum scale $p= k/a_0$ over
the cut-off scale $M$. This necessarily leads to initial states
that break the (approximate) scale invariance of the CMB spectrum,
i.e. for every comoving momentum $k$ mode the initial state
correction is slightly different, simply because they correspond
to different physical momenta at the initial time $t_0$. Clearly,
this is an example of the type (b) boundary conditions discussed
in the last section.

By contrast, this generic breaking of scale invariance in boundary
EFT differs from the type (a) approaches in which bulk modes are
treated identically by imposing an initial condition, without
explicit momentum dependence, for all modes at some fixed physical
cut-off scale $M_{cutoff}$. Momentum dependence is only implicitly
allowed through dependence on the background geometry, i.e.
through a time-varying $H$. In this framework, therefore, one
enforces the breaking of scale invariance via the slow-roll
behaviour of the background, which itself breaks de Sitter scale
invariance in the bulk. Hence, this approach also preserves
near-scale invariance of the spectrum of perturbations.

It is worth emphasizing that whereas the boundary effective field
theory method introduces a spacelike hypersurface ($t=t_0$) in
spacetime on which boundary conditions are specified, in the
approach just described, boundary conditions are specified on a
hypersurface in energy-momentum space, $E = M_{cutoff}$, which can
be referred to as the New Physics Hypersurface (NPH). Notice too
that the NPH approach does not conflict with boundary EFT per se
(one can always evolve/devolve boundary conditions specified at
different times on the NPH, to one chosen time $t_0$), but it will
not conform to generic predictions from a boundary EFT point of
view due to the {\em special} requirement of near-scale invariant
initial conditions. The EFT and NPH methods can thus be said to
represent two separate classes of boundary
conditions\footnote{Nearly all known examples
  of Planck scale modifications to the CMB fall into the two classes of modifications we have
  discussed. We will therefore limit our attention to them.}. General
(observational) consequences of initial state modifications in this
class have been described in \cite{generic-NPH}.



In light of these formal considerations of both the expectation and
relevance of initial state effects in the CMB, the pressing question
is how initial state effects alter the standard predictions based on
the Bunch-Davies state. Given a basis $u_k,u^*_k$ for the two
linearly independent solutions to the wave equation in the
inflationary background spacetime, the initial conditions determine
a unique linear combination
\begin{eqnarray}
\label{1a}
v_k &=& N(k) \left[ u_k + b(k) u_k^* \right] \, , \nonumber \\
v_k^* &=& N(k)^* \left[ u_k^* + b(k)^* u_k \right] \, .
\end{eqnarray}
Klein-Gordon normalization of the mode functions $v_k$ implies that
$|N(k)|^2 = \frac{1}{1-|b(k)|^2}$. The power spectrum of
perturbations is proportional to the absolute value $P(k) \propto
|v(k)|^2$. The (complex) parameter $b$ is known as the Bogoliubov
parameter, and we shall follow the convention that the standard
Bunch-Davies choice of initial conditions corresponds to $b=0$.
Compared to the standard Bunch-Davies form one thus obtains for the
power spectrum (defining the phase $\delta$ through $u_k =
e^{i\delta} \, |u_k|$)
\begin{eqnarray}
\label{2a} P(k) \propto \frac{1}{1-|b(k)|^2} && \left[ \left( 1 +
|b(k)|^2 + e^{2i\delta} b(k)^* \right. \right. \nonumber \\
&& \left. \left. + e^{-2i\delta} b(k) \right) \,
|u_k|^2 \right] \, .
\end{eqnarray}
Since the spectrum is evaluated for modes $p> H$ we know that
the phase $\delta$ is $k$-independent and therefore just corresponds to
an overall phase. Assuming that the corrections 
are small, i.e. $|b(k)| \ll 1$, the final expression for small initial state
modifications to the (BD) primordial spectrum of inflationary
perturbations is
\begin{equation}
\label{3a} P(k)\approx P_{BD}(k) \Big[ 1 + 2|b(k)|\cos (\alp(k) +
\delta) \Big] \, .
\end{equation}
The {\em distinctive} feature of generic initial state modifications
is thus the appearance of an oscillatory signal on top of the
standard BD spectrum with the period and amplitude determined by the
complex Bogoliubov parameter $b(k)=|b(k)|\exp(i\alp(k))$. Throughout
the rest of this proceeding we will drop the appearance of (arbitrary)
constant phases $\delta$.

\subsection{Corrections to the primordial spectrum from
  scale-invariant initial conditions}

The above expression for the corrections to the power spectrum
directly shows the effect of near scale-invariant initial
conditions. They correspond to explicitly $k$-independent Bogoliubov
parameters $b$, though they may have implicit $k$-dependence
through the background value of the Hubble parameter $H$. In a
pure de Sitter background with constant $H$ the scale invariance
is exact. In scenarios where the size of the Bogoliubov parameter
is tied to the New Physics Hypersurface where $p(t)= M =
M_{cutoff}$, the minimal choice (i.e. the minimal
uncertainty/'empty' state at the NPH)
 is $b=\frac{H}{2iM}\,
e^{-2iM/{H(1-\eps_H)}}$ with $\eps_H$ the (Hubble) slow roll
parameter of the inflationary background \cite{generic-NPH}. Any
$k$-dependence in these near-scale invariant scenarios is induced
by the time dependence --- and therefore $k$ dependence --- in the
Hubble parameter $H$. For a quasi-de Sitter background $H$ depends
on the momentum scale as $H \propto k^{-\eps_H}$

There is some reason to believe that these New Physics Hypersurface
scenarios, with a generalized Bogoliubov parameter $b =
\tbeta \frac{H}{2iM} e^{-2i\frac{M}{H}}$, are the only consistent
scale invariant modifications to the Bunch-Davies initial
state\footnote{One needs the exponential factor to avoid
  non-localities at order $H$ \cite{BEFT}. Any subleading
  prefactor will be unobservable.}. The power spectrum in this
consistent subclass is described by the expression
\begin{eqnarray}
  \label{eq:3}
  P(k) \approx P_{BD}(k)\left[1+ {\tbeta}\frac{H(k)}{M}\sin
  \left(\frac{M}{H(k)}\right) \right]
\end{eqnarray}
We will consider this case only from now on and compare it to the
generic predictions made by the boundary EFT formalism.

\subsection{Corrections to the primordial spectrum from boundary EFT}

In the boundary EFT formalism one finds instead that the amplitude
and phase of $b$ are $k$-dependent functions. This requires a bit
more explanation (for all the details we refer to \cite{BEFT}, see 
also the related work \cite{CH}),
because the Bogoliubov parameter $b(k)$ is not a natural parameter
in the effective action. The starting point in this case is the
boundary action,
\begin{equation}
\label{4b} S_B = \int_{t=t_0} d^3x \sqrt{\tilde{g}} \left(
-\frac{1}{2} \kappa_{BD} \phi^2 \right)~,
\end{equation}
introduced at some initial time or scale factor $a_0(t_0)$. Using
the machinery of effective field theory, starting with a bare
coupling reproducing the Bunch-Davies initial state, one can
calculate corrections to this bare coupling $\kappa_{BD}$ by
considering the effect of higher-derivative (irrelevant) operators
in the boundary theory. The assumption of new physics at some
physical cut-off scale $M$ --- close to the Planck scale ---
naturally introduces these irrelevant operators. They encode the
particulars of the unknown high energy physics order by order in an
expansion in the physical momentum $p_0 = \frac{k}{a_0}$ over the
cut-off scale $M$. On the basis of straightforward dimensional
analysis we {\it generically} expect the leading correction to the
bare Bunch-Davies coupling constant $\kappa_{BD}$ to be of the form
(note that $\kappa$ has dimensions of mass)
\begin{equation}
\label{4a} \kappa(k) \approx \kappa_{BD} + \beta \left(
\frac{k^2}{a_0^2 M} \right) \, ,
\end{equation}
Under the assumption of naturalness the coefficient $\beta$ is
moreover expected to be of order $1$  (although, it is
entirely possible that $\beta$ is fine-tuned in the real world).

To connect with the general power spectrum expression (\ref{3a}), we
translate this generic boundary EFT correction to an expression for
the Bogoliubov parameter $b(k)$. To do so, we remind ourselves that
the boundary action was introduced to set the initial condition.
Varying the action, one finds that the coupling $\kappa$ corresponds
to the following boundary condition on the scalar inflaton field
$\phi$
\begin{equation}
\label{5a} \left. \partial_n \phi \right|_{a_0}= - \kappa
\,\phi(a_0) \, ,
\end{equation}
where $\partial_n = H \frac{\partial}{\partial \ln a}$ corresponds
to the normal derivative with respect to the boundary. From
(\ref{5a}) it is straightforward to deduce a relation between the
coupling $\kappa$ and the Bogoliubov parameter $b$. Expand the
scalar field in a basis of two independent mode functions, allowing
for an arbitrary Bogoliubov rotation, and substitute this into
(\ref{5a}) to obtain
\begin{equation}
\label{6a} b(k) = - { \kappa(k) u_{k}(t_0) + \left. \partial_n u_k
\right|_{t=t_0} \over \kappa(k) u^*_{k}(t_0) + \left. \partial_n
u^*_k \right|_{t=t_0}} \, .
\end{equation}
This equation relates $b(k)$ and $\kappa(k)$ in general. What we are
really interested in is a relation between the Bogoliubov parameter
$b$, as defined with respect to the BD mode functions, and the
leading irrelevant correction to the bare BD coupling $\kappa_{BD}$
(\ref{4a}). Expanding (\ref{6a}) to leading order in corrections to
the BD state, using the BD mode functions $u_k$ and the
normalization conditions, we get that
\begin{equation}
\label{7a} b(k) = i a_0^3 \, (u_{k}(t_0))^2 \, \beta \left(
\frac{k^2}{a_0^2 M} \right)+ \ldots \, .
\end{equation}

Now we can use (\ref{3a}) to evaluate the effect of the leading
higher derivative correction in boundary EFT to the initial
conditions on the primordial inflationary power spectrum. The
explicit BD mode functions (for a massless scalar field) will differ
depending on the specific inflationary background. The limit where
the comoving momentum $k$ is much larger than the comoving horizon
size {\em at the initial time} $t_0$, i.e. when  $k \gg a_0H$,  is
universal, however. For all inflationary backgrounds the $y_0 \equiv
k/a_0H \gg 1$  corrections to the power spectrum are
\begin{equation}
\label{9a} P(k) \approx P_{BD} (k) \left[ 1 + \beta \frac{k}{a_0 M}
\sin (2 y_0) \right] \, .
\end{equation}
Notice the presence of {\em two}
relevant scales in this expression: $k_{H}=a_0H$ and the `comoving
cut-off scale at the initial time' $k_M \equiv a_0 M$. One might
take issue with this introduction of a second scale $1/\eta_0$. In a
most conservative scenario one can think of it as the beginning of
inflation or the 'Planck time' before which GR breaks down. We will
elaborate on the interpretation and theoretical expectation of the
(period) scale $k_H$ and the (amplitude) scale $k_M$ in the next
subsection.

\subsection{Observable parameters and physical quantities}

As explained and emphasized, it is a generic feature that initial state corrections
are characterized by oscillations on top of the standard spectrum of
fluctuations. This implies that in principle there will be two, a
priori, independent observable parameters extractable from (future)
CMB data; the amplitude and the period of an oscillatory component
of the primordial power spectrum\footnote{For very high comoving
momentum modes this can  directly be extrapolated to an amplitude
and period in the observed multipole moments $C_l$ to reasonable
approximation; although in truth a full deconvolution is called
for \cite{Pheno}.}. Preliminary data extraction studies have indicated
that these oscillatory features are indeed expected to be decipherable
in future CMB experiments (under optimistic assumptions for the ratio
$H/M$) \cite{Pheno}. The distinction between the generic boundary EFT
prediction and the near-scale invariant NPH proposal for initial
state corrections is in the $k$-dependence of these two observable
parameters.

For the boundary EFT prediction, the qualitative behaviour of the
corrections depends crucially on the relative value of the scale
$k_H$ and $k_M$ with respect to the range of comoving momentum modes
present in the observable CMB, $k \in [k_{min}, k_{max}]$. As the
scale $k_M$ corresponds to the comoving cut-off scale, beyond which
the boundary EFT formalism breaks down, we {\em must} require that
$k_{max} < k_M$. Now, the ratio between the period and the cut-off
is a physical quantity given by $\frac{k_H}{k_M} = \frac{H}{M}$ 
\footnote{\label{conundrum} This
dimensional relation between $k_H$ and $k_M$ does yield a conundrum.
Since the boundary action should preferentially be introduced before
the time the lowest modes in the CMB were formed, we expect $k_H \lesssim
k_{min}$. In other words the scale $k_H$ is expected to be at the
least the horizon size of the lowest mode in the CMB at $t=t_0$.
This simple observation creates an unexpected tension between 1) the
physical idea that modes become non-dynamical when they exit the
horizon, 2) the wish to describe the full four-orders of magnitude
range of CMB modes with one boundary EFT with fixed time initial
conditions and 3) the expectation that $H/M$ is of order $10^{-2}$.
Because the power spectrum follows from linear analysis where modes
do not interact 1) can be sacrificed without loosing its inherent
idea. For an interacting field theory it remains an open question,
however, how to resolve the tension between 1), 2) and 3).}.  Thus we
see that within a consistent boundary EFT description there is a
lower bound on the period
\begin{equation}
\label{10a} k_H \gtrsim \left( \frac{H}{M} \right) \, k_{max} \, .
\end{equation}
This theoretical estimate is important because the scale $k_H$ sets
the period of the oscillations in the spectrum, which can be read
off from (\ref{9a}) to equal
\begin{equation}
\label{11a} \Delta k = \pi k_H \gtrsim \left( \frac{H}{M} \right) \,
\pi \, k_{max} \, .
\end{equation}
Extrapolating these constraints to constraints in multi-pole space
$l$, i.e. $\Delta l = \pi l_H \gtrsim \left(\frac{H}{M} \right) \,
\pi \,l_{max}$, we can deduce a lower bound on $H/M$ beyond which
the oscillations are too frequent and are washed out of the data.
Since we know that the current $\pi l_{max} \lesssim 10^4$ and
assuming that a period $\Delta l \gtrsim 10$ is observable, we find
that $H/M \gtrsim 10^{-3}$ for oscillations to be detectable in the
CMB. If $H/M$ is at the one percent level, one would expect to see
oscillations with an estimated period around $\Delta l \sim 100$.

Gratifyingly, it is also for values of $H/M \gtrsim 10^{-3}$ that
the amplitude of the signal is at the same order of or larger than
the inherent cosmic variance ambiguity in the CMB. For a period of
order \mbox{$\Delta l \sim 10$} the larger part of the observable CMB
spectrum ($l_H < l \leq l_{max}$) is well approximated by
({\ref{9a}). From the theoretical and detectability constraints
discussed above, one finds that the amplitude $A(k)=\cA\,k$, with
$k_{H} \leq k \leq k_{max}$, runs between
\begin{equation}
\label{11b} \beta \left( \frac{H}{M} \right) \leq \cA\, k \leq \beta
\, .
\end{equation}
This is easily beyond the $1\%$ cosmic variance level around $k \sim
k_{max}$, unless $\beta$ is fine-tuned and unnaturally small. Here
we should also point out that the observed near scale invariance of
the CMB spectrum could a priori significantly constrain $\beta$
as $k$ approaches $k_{max}$. However, as it turns out, and mainly due to
the oscillatory nature of the correction, this does not lead to a
severe constraint on $\beta$. By dividing the observable parameters
of the oscillations one would probe the scale of new physics
directly
\begin{equation}
\label{12a} \frac{\Delta l_{obs}}{\cA_{obs}} = \beta \left(
\frac{H}{M} \right) \, . \end{equation}
Under the assumption of
naturalness ($\beta \approx 1$) this fixes the interesting ratio of
scales $H/M$. Moreover, if tensor modes are observed, the Hubble
scale $H$ will be known independently. The presence of CMB
oscillations with a constant period in $k$ or $l$ would then allow a
determination of the scale of new physics $M$ through the boundary
EFT formalism (again assuming naturalness).

It is a qualitative difference in the periodicity of the
oscillations that distinguishes the near-scale invariant New Physics
Hypersurface proposal. Whereas the generic prediction from boundary
EFT was a constant period in $k$, the NPH proposal yields
oscillations with a constant period in $\ln (k/k_{pivot})$. Here
$k_{pivot}$ is the arbitrary pivot point in $k$ space where the
normalization of the observed power-spectrum is set and compared to
which slow roll is measured (e.g. COBE used $k_{pivot}=
7.5H_{present}$.). Specifically the periodicity is given by
\begin{equation}
\label{13a}
\Delta \ln \frac{k}{k_{pivot}} = \frac{\pi H_{pivot}}{M\eps_H} \, .
\end{equation}
This allows us to deduce
how many oscillations we expect in the spectrum. Current CMB
measurements range from roughly $ 10^{-4}H_{present} \leq k \leq
H_{present}$ or
\begin{equation}
\label{13b}
-4\ln 10 -\ln \frac{k_{pivot}}{H_{present}} \leq
\ln k \leq -\ln \frac{k_{pivot}}{H_{present}} \, .
\end{equation}
Therefore the number of full oscillatory periods present in the CMB
ought to be
\begin{equation}
\label{13c}
N= 4\ln 10 \frac{M\eps_H}{\pi H}  \simeq \frac{3 M\eps_H}{H} \, .
\end{equation}
For the observed estimate of
$\eps_H \leq 0.01$ and the optimistic scenario that $M/H \sim 10^2$
we expect to see 1-10 oscillations over the whole power spectrum
(see e.g. figure 1 in \cite{generic-NPH}).

An advantage of the near-scale invariant NPH proposal is that it
allows one to determine the ratio of scales directly from the
period of these oscillations. This is provided that the slow roll
parameter $\eps_H$ is known. No appeal to naturalness is needed.
In fact with the knowledge of the ratio of scales we can test
directly the deviation $\tbeta$ from the standard Bunch-Davies
state. A significant difference from unity for this number could
be interpreted as an element of fine tuning at work.

Another important distinction between the NPH and EFT scenarios is 
that the EFT corrections grow with increasing $k$, whereas the NPH modifications 
decrease with increasing $k$. The physics behind this is clear: in EFT the effects 
grow larger as you approach the cut-off scale. In NPH the $k$-dependent 
corrections are proportional to the Hubble scale, which decreases in time; 
larger $k$ modes exit the horizon later and are therefore affected by 
a smaller Hubble scale. This indeed confirms that the natural place to look 
for corrections in the NPH scenario is the CMB spectrum. At the same 
time however, it suggests that the more natural place to look for EFT 
corrections would instead be in large $k$ power spectra that seed  
galaxy formation.

\renewcommand{\arraystretch}{2}

\begin{table*}[htbp]
  \centering
  \begin{tabular}{|c||c|c|}
   \hline
 & Boundary EFT & Minimal NPH \\
\hline \hline Power Spectrum & ~~~${\displaystyle P=P_{BD}\left(1+\cA k \sin
\left( \frac{2 \pi k}{\cC} \right)\right)}$~~~ &~~~ ${\displaystyle P=P_{BD}\left(1+{\sf A}
\sin\left(\frac{2\pi}{\sf C} \ln \frac{k}{k_{pivot}}\right)\right)}$~~~
\\[.07in]
\hline \hline Amplitude &  ${\displaystyle \cA =\beta\frac{1}{a_0M}}$ &
  ${\displaystyle  {\sf A} ={\tbeta}\frac{H_{pivot}}{M}}$ \\[.07in]
\hline Period & $ {\displaystyle \Delta k = {\cC} = \pi a_0 H}$ &
  ${\displaystyle \Delta \ln\frac{
k}{k_{pivot}} = {\sf C} = \frac{\pi H_{pivot}}{M \eps_H}}$ \\[.07in]
\hline Number of Osc.   & ${\displaystyle N \lesssim
  \frac{M}{\pi \, H}}$ &
${\displaystyle N \simeq \eps_H \frac{M}{\pi \, H} \ln
  \frac{k_{max}}{k_{min}}}$
\\[0.07in]
\hline \hline Ratio of Scales &   ${\displaystyle {\cA} \cdot \Delta k = \beta
  \frac{H}{M} }$ &  ${\displaystyle {\sf A} = \tbeta\frac{H}{M}~,~\frac{\eps_H {\sf C}}{\pi}=
\frac{H_{pivot}}{M}}$\\[.07in]
\hline
  \end{tabular}
  \caption{Phenomenological signatures of initial state effects on the
  primordial CMB power spectrum. $P_{BD}$ is the Power Spectrum
  computed w.r.t. Bunch-Davies initial conditions and an arbitrary constant phase has been dropped.}
  \label{tab:1}
\end{table*}

\section{Conclusions}

In table \ref{tab:1}, we have summarized our discussion of initial
state effects in the CMB that arise within boundary Effective
Field Theory at a fixed time and within the special class of
near-scale invariant New Physics Hypersurface initial conditions.
The qualitative difference between the two scenarios is clearly
the behaviour of the periodicity in $k$-space: constant for
generic initial conditions in boundary Effective Field Theory;
logarithmic for near-scale invariant New Physics Hypersurface
scenarios. A secondary and related aspect is a linearly growing
amplitude for boundary Effective Field Theory but a constant
amplitude for near-scale invariant New Physics Hypersurface
proposals. Note that under the assumption of naturalness the
boundary EFT formalism predicts a marginally bigger window of
opportunity in $H/M$ space as compared to the NPH scenarios.

Throughout our discussion, we have taken a decidedly
phenomenological perspective on transplanckian physics,
emphasizing --- as in table \ref{tab:1} --- the generic signatures one
would hope to find if high energy physics does in fact yield an
observational imprint on the CMB. For completeness, we briefly
note one important theoretical issue. Part of the growing
literature on transplanckian physics has involved a debate about
the expected magnitude of transplanckian corrections. In
\cite{egks,MNPH} it was argued that
--- as in the approaches we have reviewed above --- we should
expect order $H/M_{cutoff}$ corrections, a conclusion borne out by
many explicit studies \cite{Brandenberger,Others}. However, in
\cite{Kaloper} it was argued that corrections could at most be of
order $(H/M_{cutoff})^2$. Due to cosmic variance limitations, this
constitutes a qualitative difference as to whether the corrections
can, even in principle, ever be seen. The disparity between these
two claims arose because \cite{egks,MNPH} explicitly
--- and \cite{Brandenberger, Others} implicitly --- allowed for
modified boundary conditions and modified dynamics whereas
\cite{Kaloper} only allowed for modified dynamics (coming from
higher order operators in the \emph{bulk} effective field theory).
The second paper of \cite{Kaloper} went further and argued that it
was physically inconsistent, or at the very least technically
unnatural due to large backreaction, to have any but the standard BD
boundary conditions. The advantage of working in a boundary EFT
formalism, as in \cite{BEFT}, is that backreaction can be
systematically analysed. And as shown in \cite{BEFT} and
\cite{Porrati}, large backreaction within the context of effective
field theory can be avoided, giving us a self-consistent framework
that predicts order $H/M_{cutoff}$
corrections.\footnote{Backreaction in the context of
  the NPH proposal was discussed in \cite{back2}, with qualitatively
  the same conclusion that a clear window of opportunity to detect new
  physics in the CMB remains.}

Thus, the most important observation is that with minimal
assumptions, i.e. that $H/M$ be large enough compared to errors
due to cosmic variance, the {\em generic oscillatory}
characteristics of initial state effects should be visible in
future if not current CMB experiments \cite{WMAP}. Nature could of
course have arranged itself such that the controlling coefficients
$\beta$ or ${\tbeta}$ in the Effective Field Theory or New Physics
Hypersurface scenarios --- describing the particulars of new
physics at the cutoff scale $M$
--- are small. This would be yet another fine-tuning to bewilder us
theorists. Absent that, with guarded optimism we can imagine that in
the not too distant future we might catch the very first
experimental glimpse of near-Planck scale physics.

\bigskip

{\bf \noindent Note added:}\\

As indicated at the end of section 2, a more natural place to look
for EFT effects might be the power spectrum of density perturbations deduced 
from galaxy surveys (like for instance the Sloan Digital Sky Survey \cite{SDSS}). 
We would like to thank G. Holder, W. Kinney, R. Easther and L. Hui for pointing out to us
that over a large range of length scales the galaxy survey power spectrum is still 
governed by linear physics. This potentially allows for new physics signatures according 
to the EFT formalism. However, assuming the comoving cut off scale $k_M$ is high 
enough to cover the galaxy survey scales does exacerbate the conundrum mentioned 
in footnote (\ref{conundrum}).

\bigskip 
\begin{acknowledgments}
BRG would like to thank the organizers of the Texas at Stanford
conference for a lively and stimulating conference. BRG, JPvdS and
KS acknowledge financial support from DOE grant
DE-FG-02-92ER40699. The work of GS is supported  in part by NSF
CAREER Award No. PHY-0348093, DOE grant DE-FG-02-95ER40896 and a
Research Innovation Award from Research Corporation. The authors
gratefully acknowledge support from the Ohrstrom Foundation.

\end{acknowledgments}

\bigskip 

\end{document}